\documentclass[letter,12pt]{article}
\usepackage{graphicx,amssymb,amsmath,epstopdf,color,hyperref,subfig}

\def\beq{\begin{equation}}
\def\eeq{\end{equation}}
\def\bea{\begin{eqnarray}}
\def\eea{\end{eqnarray}}

\def\mev{\, {\rm MeV}}
\def\gev{\, {\rm GeV}}

\newcommand{\gsim}{\lower.7ex\hbox{$\;\stackrel{\textstyle>}{\sim}\;$}}
\newcommand{\lsim}{\lower.7ex\hbox{$\;\stackrel{\textstyle<}{\sim}\;$}}

\begin{document}

\begin{flushright}
\today
\end{flushright}

\vspace{0.07in}

\noindent
\begin{center}

{\bf\large Higgs boson decays into narrow di-photon jets \\ and their search strategies at the Large Hadron Collider}

\vspace{0.5cm}
{Benjamin Sheff, Noah Steinberg, James D. Wells}

{\it Leinweber Center for Theoretical Physics \\
Physics Department, University of Michigan \\
Ann Arbor, MI 48109-1040 USA}\\

\end{center}

\noindent
{\it Abstract:} 

In many extensions of the Standard Model the Higgs boson can decay into two light scalars each of which then subsequently decay into two photons. The underlying event is $h\to 4\gamma$, but the kinematics from boosted light scalar decays combined with realistic detector resolutions may fail to register the events in straightforward categories and thus may be lost. In this article we investigate the phase space for highly boosted di-photon events from these exotic Higgs decays and discuss search strategies that aim to capture and label events in this difficult region. In the process we develop a new category, $\xi$-jets,  which identifies with high selectivity highly collimated di-photon decay modes of the Higgs boson.

\vfill\eject

\tableofcontents

\section{Introduction}

Nearly a decade after the discovery of the Higgs boson it remains to be decided whether the discovered particle interacts with other known elementary particles in precisely the way the Standard Model dictates\cite{Sirunyan:2018owy,Aaboud:2018xdt,Belanger:2013xza,Aad:2019mbh,Khachatryan:2016vau,Curtin:2013fra}. Deviations from SM expectations can arise by virtue of the Higgs boson being composite, part of a larger Higgs sector, coupled through its portal interactions to hidden sector states, or embedded in extra dimensions to name just a few examples. Alternatives remain viable because the SM Higgs boson couplings to other SM states are known only to at best 10\% for some, and only to within ${\cal O}(1)$ factors for others, including muon, electrons, charm, and Higgs self interactions\cite{Aad:2020xfq,Aad:2019uzh}. The possibility of the Higgs boson decaying into final states that are not allowed by the SM is also not constrained well in many cases. 

In this article we take up the case of the Higgs boson ($h$) decaying into other very light scalars ($\phi_1$ and $\phi_2$) where each subsequently decays into photons, 
\bea
h\to\phi_1\phi_2\to (\gamma\gamma)(\gamma\gamma)~~~{\rm (target~observable)}.
\label{eq:observable}
\eea
In several different limits this process has been studied already\cite{Dobrescu:2000jt,Draper:2012xt}. In the case of $\phi_{1,2}$ both having mass above about $10\gev$ one finds that the events register as unambiguous $4\gamma$ events in the detector that can be searched for well. Within this regime, current studies limit this process to $B(h\to 4\gamma)\lesssim 3\times 10^{-4}$~\cite{atlas_4_photon_search,Chang:2006bw}. 

On the other extreme, if $\phi_{1,2}$ both have mass less than a few hundred MeV, the photons from $\phi_i\to\gamma\gamma$ are so collimated coming from the highly boosted $\phi_i$ resultant from their parent Higgs decay, that each $\phi_i$ decay appears to go to a single photon. In that case, $h\to\phi_1\phi
_2$ is simply combined with the standard $h\to\gamma\gamma$ analysis, and it becomes a statistical question to determine what overabundance of such a signal would be consistent with data. At 95\% CL the answer to this question is that the branching fraction of non-SM contributions to $B(h\to\gamma\gamma)$ cannot exceed $2.2\times10^{-4}$\cite{PDG}. Such light scalars may also be disentangled from the SM $h\to\gamma\gamma$ process with sophisticated substructure techniques\cite{photon_jets,ATLAS:2012soa}.

Combining both extremes leads to an apparent detection $h\to 3\gamma$. This arises when one of the $\phi_i$ has mass less than a few hundred MeV and the other more than about $10\gev$. This process is forbidden in the SM, and the branching ratio is currently limited to $B(h\to 3\gamma)\lesssim \times 10^{-3}$, as can be gleaned from~\cite{chawdhry_nnlo_2020}.

In between these two extremes, from the point of view of observables, is a murky region where the mass of one or both $\phi_i$ states is between $\sim 0.1\gev$ and $10\gev$. In that case, the two photons coming out of the $\phi_i$ decays are not highly collimated nor or they cleanly separated. Roughly speaking, the ATLAS and CMS detectors see something distinct from a standard photon but that also does not register as two photons when the photon separation is between $0.04<\Delta R<0.4$\cite{Aaboud:2018djx,atlas_photon_efficiency}.  It is this difficult middle ground region that we wish to address in this letter.

It should be stated that extending the scalar sector of the SM by one (or multiple) singlets is a mature and well studied subfield~\cite{Robens:2015gla,Barger:2007im,Robens:2019kga}. Much of the parameter space for exotic heavy and light  scalars (relative to the Higgs boson mass) is well constrained by direct searches and by precision electroweak measurements~\cite{Lopez-Val:2014jva}. Our simplified model highlights a region of parameter space in a class of singlet extended models that has been less explored by previous studies.

The value in exploring such a regime lies in its ability to utilize the available experimental power from the LHC to investigate one of the most interesting loose ends in the Standard Model. Many models exist coupling new light scalars to the Standard Model in ways that are highly susceptible to the search strategy we advocate here~\cite{Csaki:2020zqz}. The nature of the Higgs boson makes such couplings to new physics generic and apparent in a broad swath of theory parameter space. Furthermore, the rough knowledge we have of the Higgs boson to date deserves significant tightening in every reasonable direction. Our goal here is to consider this particular case in detail, highlight the experimental challenges for discovery, proffer some suggestions, and suggest a benchmark theory with points that may be useful for serious further study by experimental groups within the ATLAS and CMS collaborations.

\section{Theory description}

The phenomenon we are after is $h\to\phi_1\phi_2$ with subsequent decay of $\phi_i\to\gamma\gamma$. Such decays arise generically in a broad class of BSM theories, many of which give rise to additional exotic phenomena. Most commonly these are other, similar gauge interactions, such as $Z\to\phi\gamma$, but the possibilities are wide and varied. Many BSM theories of this type are not yet constrained by experiment and have their most accessible phenomenon as $h\to\phi_1\phi_2\to 4\gamma$, if there are dedicated searches for it. Our focus lies in this last type of theory.

To devise an experimental strategy and analysis to discover this class of targeted theories, we must begin by constructing a representative theory within the class and finding ways to find evidence for it. Ideally the representative theory should be maximally simple without losing the key features under consideration for our exotic Higgs decays. In this case, there is such a simple theory, and its lagrangian is
\bea
{\cal L}& = &{\cal L}_{\rm SM}+\frac{1}{2}(\partial_\mu\phi_1)(\partial^\mu\phi_1)+\frac{1}{2}(\partial_\mu\phi_2)(\partial^\mu\phi_2)-\frac{1}{2}m^2_1\phi^2_1-\frac{1}{2}m^2_2\phi^2_2 \nonumber \\
& & +\lambda_\phi |H|^2\phi_1\phi_2 
 +\frac{1}{\Lambda_1}\phi_1 F_{\mu\nu}F^{\mu\nu}+\frac{1}{\Lambda_2}\phi_2 F_{\mu\nu}F^{\mu\nu}~~{\rm (representative~theory)}
\label{eq:theory}
\eea
where $F^{\mu\nu}$ is the photon field strength tensor. Of course, one could write down non-trivial $|H|^2\phi_1^2$ and $|H|^2\phi_2^2$ terms among others, but that would which add complexity without contributing significantly to the final phenomenology. One might also object that $\phi_i F_{\mu\nu}F^{\mu\nu}$ should be traded in for gauge-invariant couplings of $\phi_i$ to hypercharge field strength tensor $\phi_i B_{\mu\nu}B^{\mu\nu}$ and $SU(2)$ field strength tensor $\phi_i W^{a}_{\mu\nu}W^{a,\mu\nu}$. That would be fine, except that upon diagonalizing these interactions to those of the mass eigenstates one finds nevertheless $\phi_i F^2$ terms, which will completely dominate in the decays of $\phi_i$ over $\phi_{i}Z^{\mu\nu}F_{\mu\nu}$ and $\phi_i Z^2$ terms due to the $Z$ boson being much heavier than the $\phi_i$ that we will consider below.\footnote{Barring any tuned cancellations, typical branching ratios to $\gamma\gamma$ are $10^7(10^{15})$ times larger than the branching ratios to $\gamma Z^{*}(Z^{*}Z^{*})$} The $\phi_iZF$ interaction can give rise to $Z\to \phi_i\gamma$ decays, constrained by searches at the Tevatron and the LHC\cite{Aaltonen:2013mfa,atlas_4_photon_search}, but as the scale of $\Lambda_i$ becomes higher, this constraint goes away while $B(\phi_i\to\gamma\gamma)$ remains 100\%\footnote{As the $\Lambda_{i}$ increase, so does the decay length of the scalar. We have checked that the scalar decay length can be under $1\,{\rm mm}$ even for large (PeV)  values of $\Lambda_{i}$ which evade the  $Z\to \phi_i\gamma$ constraint.}. For that reason we drop these extra consideration and extraneous interactions from the theory description and retain only the lagrangian of Eq.~\ref{eq:theory}.

From the point of view of devising experimental search strategies to find evidence for the Higgs boson decaying into a single light scalar, say $\sigma$ such that $h\to \sigma\sigma\to 4\gamma$, the benchmark theory above is adequate. It merely corresponds to the case of $m_1=m_2$. That is not to say the two theories are exactly the same, only that the subsequent search strategies are the same. That is why we propose to work with only one theory -- the representative theory of Eq.~\ref{eq:theory} -- which we believe to form a basis upon which benchmark points can be established and strategies devised.

\section{Photon $\xi$-jets}

\begin{figure}[t] 
\begin{center}
\includegraphics[width=0.85\textwidth]{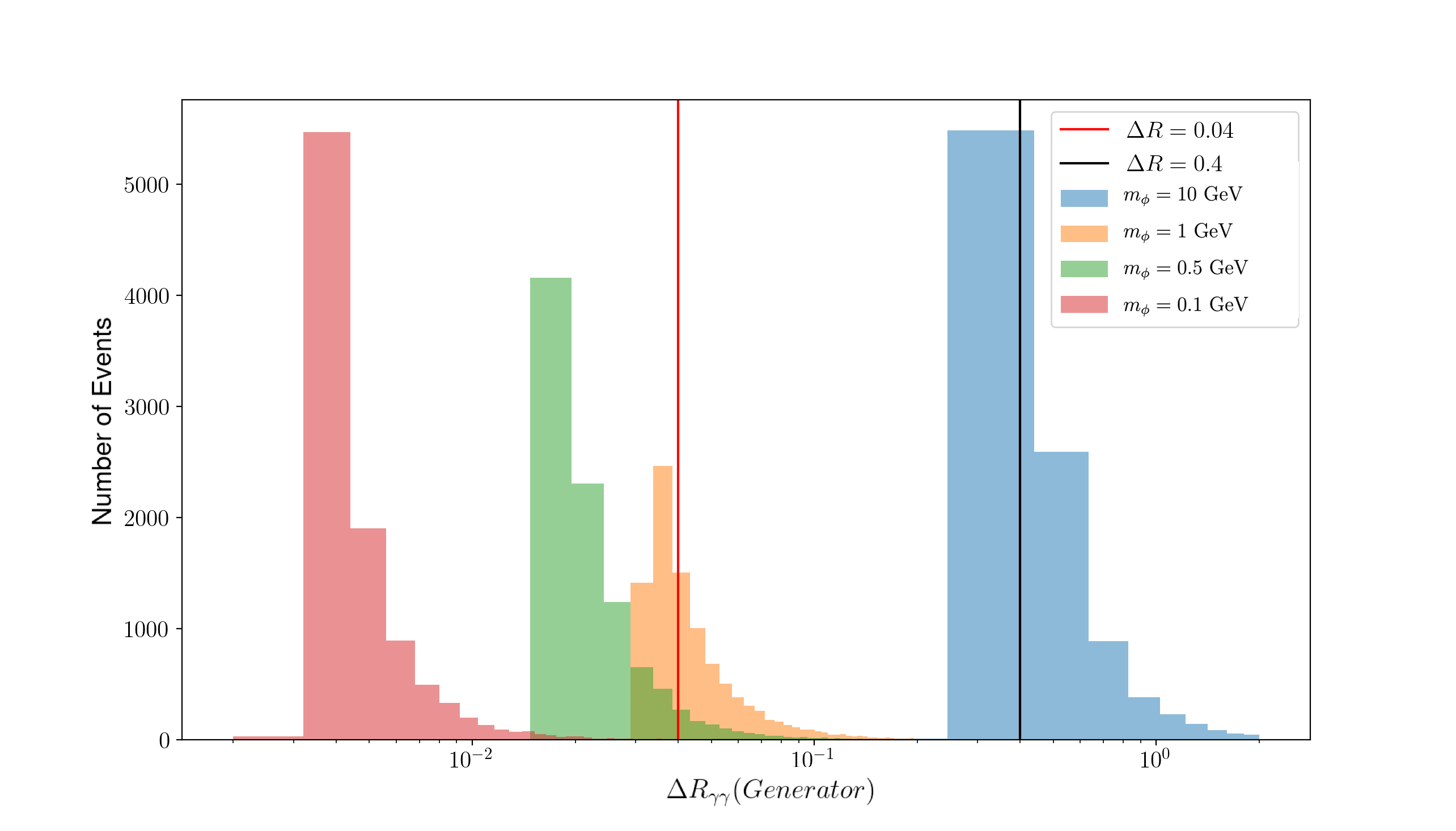} 
\caption{$\Delta R$ separation between photon pairs from $\phi_{i}$ decays, sampled over 10,000 events at varying masses of the BSM scalars.}
\label{fig:DeltaRs}
\end{center}
\end{figure}

As we mentioned in the introduction, the target observable of Eq.~\ref{eq:observable} implies photon separation from $\phi_i$ decays that is sensitive to the $\phi_i$ masses. This is illustrated in fig.~\ref{fig:DeltaRs}, which shows that $m_\phi=10\gev$ gives well separated photons ($\Delta R>0.4$) and $m_\phi=0.1\gev$ gives very collimated photons ($\Delta R<0.04$), and mass of $1\gev$ gives intermediate separation. Recall that $\Delta R=\sqrt{(\Delta \phi)^2+(\Delta\eta)^2}$, and $\Delta\phi$ is the azimuthal angle separation and $\Delta\eta$ is the pseudo-rapidity separation of the two photons in $\phi_i\to\gamma\gamma$ decay.

 Thus, it is relatively straightforward phenomenology if both $\phi_1$ and $\phi_2$ have mass greater than $10\gev$. The two states decay into well separated photons $\phi_i\to \gamma\gamma$ and the target observable becomes four well separated photons, all reconstructing the Higgs mass $m_h=m_{4\gamma}$. Such a prospect does not require further discussion here, as all the standard tools of experimental analysis to identify well isolated photons can be employed to make straight-forward searches, as have been done in\cite{atlas_4_photon_search,Chang:2006bw}.

Likewise, if $m_{\phi_i}\lsim 100\mev$ the decays chain of $h\to \phi_1\phi_2\to (\gamma\gamma)(\gamma\gamma)$ yields highly boosted $\phi_i$ light states that can decay into highly collimated photon pairs that then register in the electromagnetic calorimeter as a single photon. Once $m_{\phi_i}$ dips below $100\mev$ that rate is nearly 100\%. Thus, experimentally, for such light $\phi_i$ the target observables register in the detector as $\gamma\gamma$ events with $m_{\gamma\gamma}=m_h$, and thus contributes to the count of such non-exotic events already produced by the direct decays of $h\to\gamma\gamma$ through top and $W$ loops. The sensitivity to this possibility then becomes a statistics question of how many exotic sources of $h\to\gamma\gamma$ events can the data tolerate. As we mentioned in the introduction, that rate is approximately $2.2\times10^{-4}$\cite{PDG}. Additional discussion is not needed here.

We then turn to the more ambiguous case in which the $\phi_i$ masses fall within the ``intermediate mass" range of $0.1\gev < m_{\phi_i}<10\gev$. Within the LHC environment, the production of Higgs bosons and their subsequent decay into such scalars yields photon pairs separated by
\beq
0.04<\Delta R_{\gamma\gamma}<0.4~~~{\rm (intermediate~separation).}
\label{eq:separation}
\eeq

It is well known that photon pairs that fall within the intermediate separation range of Eq.~\ref{eq:separation} are extremely difficult to separate or identity. We will speak much more on that below, but here we wish to pay respect to that difficulty by giving it a name. We call two photons that are within the range specified by Eq.~\ref{eq:separation} a ``$\xi$-jet". The $\xi$-jet is a purely theoretical object, and it is defined by underlying ``truth data" and not with respect to any detector performance. If a photon has another photon within the intermediate separation annulus of Eq.~\ref{eq:separation}, and nothing else is within the outer ring of that annulus, then it ceases to be a photon and the two together form a $\xi$-jet. Such a concept can be generalized to more than two photons but it is of not much importance here to do that. We also specify as a theoretical object that a photon is defined to be either a single photon or two photons within $\Delta R<0.04$ of each other. 

With these theory definitions of photon and $\xi$-jet, our target observable is broken into several distinct and non-overlapping final states, depending on the masses of the $\phi_i$ intermediate states in the decay chain:
\beq
h\to\phi_1\phi_2\to 4\gamma~ \Rrightarrow ~4\gamma,~ 2\gamma,~ 3\gamma,~  \gamma\xi,~ \gamma\gamma\xi,~ 2\xi ~~~{\rm (observable~partitions)}
\label{eq:partitions}
\eeq
The first three of these observables we have already discussed. The remaining observables have not been fully explored in the literature, and we wish to consider them in more detail below.

\section{Benchmark model points}

We are interested in exploring three observables: $\gamma\xi$, $\gamma\gamma\xi$, and  $2\xi$. To do so we need benchmark points that give rise to each of these types of observables. They can be obtained rather straightforwardly from our representative theory of Eq.~\ref{eq:theory} where the masses of $\phi_1$ and $\phi_2$ are chosen to be various permutations of  the masses $0.1\gev$, $1\gev$ and $10\gev$. In particular, $m_\phi=0.1\gev$ generally always gives $\phi\to\gamma$ decays, $m_\phi=1\gev$ typically gives $\phi\to\xi$ decays, and $m_\phi=10\gev$ generally gives $\phi\to\gamma\gamma$ decays according to our definitions in the previous section. These are so far entirely defined theoretically. In the next section we will pursue more carefully how a theoretical $\xi$-jet registers in an experimental analysis.

\begin{figure}[t] 
\begin{center}
\includegraphics[width=0.85\textwidth]{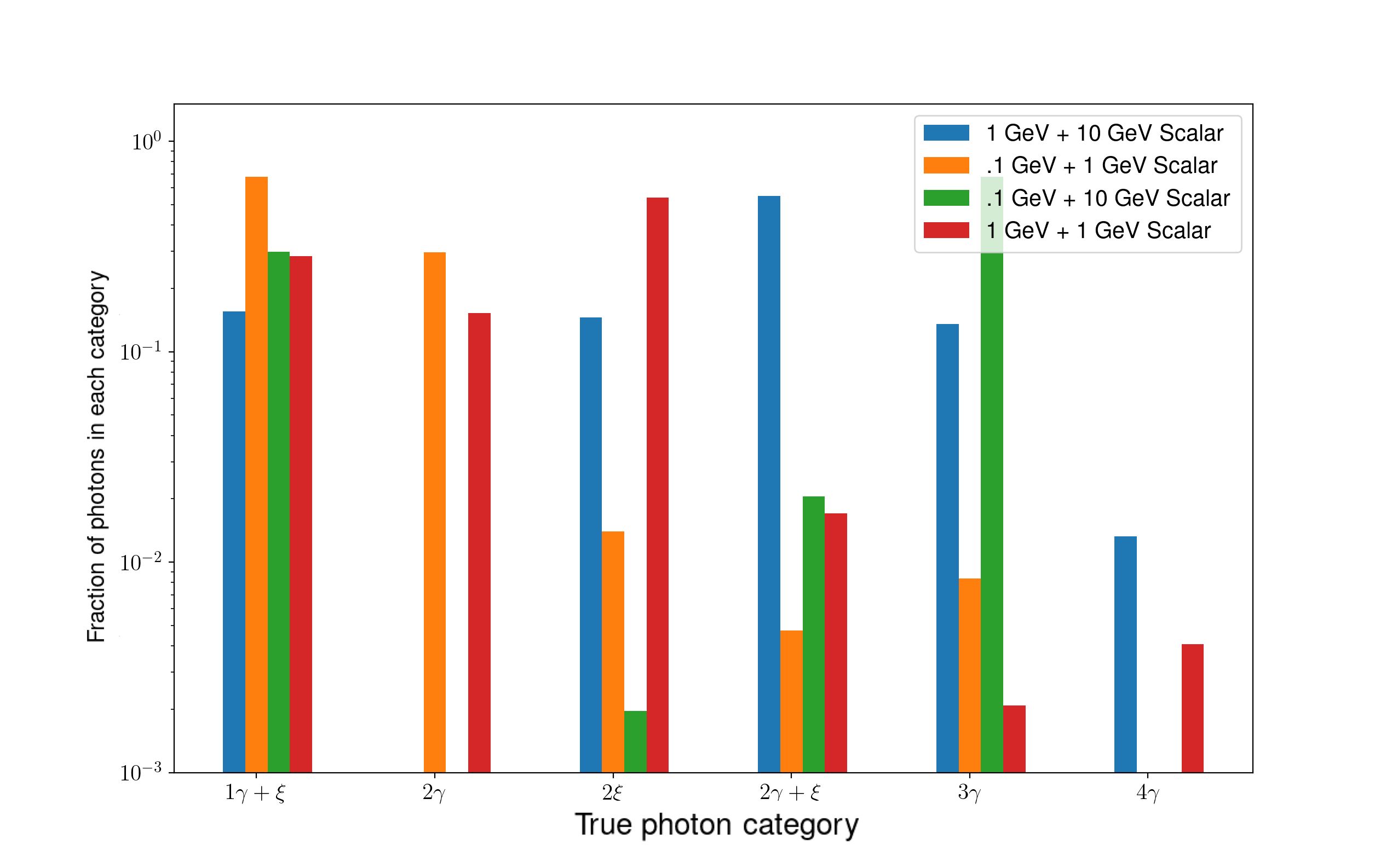} 
\caption{Branching fraction into each final state theory observable for the benchmark points A (blue), B (orange), C (red) and D (green) given in Table~\ref{table:benchmarks}.}
\label{fig:toy_model_fractions}
\end{center}
\end{figure}

From these considerations we can construct the following three benchmark points A, B, and C, specified in Table~\ref{table:benchmarks}. Fig.~\ref{fig:toy_model_fractions} shows the relative fraction of each observable for each benchmark point. The dominant and subdominant modes of decay for each benchmark point are listed in Table~\ref{table:benchmarks} and can be gleaned from the fraction data given in Fig.~\ref{fig:toy_model_fractions}. Table~\ref{table:benchmarks} shows that several combinations of light scalar masses give interesting decay signatures involving combinations of $\xi$-jets and photons which (to the authors' best knowledge) are not being searched for in current LHC analyses.

\begin{table}[t]
\centering
\begin{tabular}{ccccc}
\hline\hline
Point & $m_1$ (GeV) & $m_2$ (GeV) & Dominant mode & Subdominant mode \\
\hline
A & 1 & 10 & $\gamma\gamma\xi$ & $\gamma\xi\simeq 2\gamma\simeq 3\gamma$ \\
B & 0.1 & 1 & $\gamma\xi$ & $2\gamma$ \\
C & 1 & 1 & $2\xi$ & $\gamma\xi\simeq 2\gamma$ \\
\hline\hline
\end{tabular}
\caption{Benchmark points for $h\to\phi_1\phi_2\to 4\gamma$ which then partition into various theory-object observables (modes) according to our definitions of $\xi$ (photon pairs with $0.04<\Delta R<0.4$) and $\gamma$ (an isolated photon with $\Delta R>0.4$ or two photons within $\Delta R<0.04$). }
\label{table:benchmarks}
\end{table}

\section{Experimental search strategies}
So far our discussion has been mainly theoretical. We have identified a rare Higgs decay whose cascade we claim may be difficult to detect by experiment. In this section, we discuss how our theoretical objects translate into experimental manifestations. We have suggested that some mass ranges of $\phi_i$ are problematic for experiment. We will discuss some details on why they are challenging and some strategies by which to possibly overcome those challenges. 

\subsection{Multi-photon final states}

Isolated photons or extremely  highly collimated photons both get identified simply as photons, and analysis based on those  standard objects (photons) proceed without much subtlety regarding how to process the data into well-defined final states of $2\gamma$, $3\gamma$ and $4\gamma$. 

\subsection{$\xi$-jet final states}

Some of the final states from the decays of Eq.~\ref{eq:partitions} yield $\xi$-jets. Underneath, a $\xi$-jet is merely two photons with intermediate $\Delta R$ separation (see Eq.~\ref{eq:separation}). But a key question is, how does a $\xi$-jet, defined as a theoretical object, get processed into various experimental categories? A perfect detector would register it as merely two photons, a bad detector as a single photon or nothing, and a realistic good detector, such as ATLAS or CMS, registers it as something altogether different within several possible categories of varying sensitivity and selectivity\footnote{By perfect (non-existent) ``sensitivity" we mean a category test that passes with 100\% (0\%) rate if the underlying event is a $\xi$-jet, and by perfect (non-existent) ``selectivity"  we mean a category test that passes with 0\% (100\%) rate if the underlying event is not a $\xi$-jet. Good sensitivity means low false negative rate, and good selectivity means low false positive rate.}.

To address this question of how a $\xi$ registers in a detector it is useful to describe the various categories into which a single photon can fall. As an example we take the standard categories which ATLAS uses for photon identification. There are eight possible standard categories, six are the permutations among three isolation possibilities (non-isolated, loose isolation, and tight isolation) and two ID possibilities (loose ID and tight ID). The other two categories are jet and ``lost." Jet is the standard QCD jet from fragmentation of quarks or gluons, and ``lost" refers to the possibility that the data does not conform to any other category and is not registered in any higher abstracted category except for mere energy depositions in the detector. 

A $\xi$-jet will register with some probability into one or more of the standard photon categories. The probability to do so depends on the underlying event kinematics. Under typical assumptions, the $\xi$-jet will often register as ``lost" due to the inability to resolve the two photons yet the event covers more than one cell in the electromagnetic calorimeter which a single photon would not do. As no category becomes applicable, it has no option but to be relegated to ``lost." 

The implication of a $\xi$-jet arising from a Higgs decay being categorized as ``lost" is that an analysis that requires reconstructing the invariant mass of the Higgs boson from well-defined decay products can no longer register the events. It is therefore necessary to build a ninth category ``$\xi$-candidate" under which $\xi$ events can fall. $\xi$-candidates must be defined entirely through detector response, with the goal of producing high sensitivity to underlying $\xi$-jets with reasonably good selectivity (i.e., mostly only $\xi$-jets register as $\xi$-candidates).
          
\subsection{$\xi$-candidates}
          
A detailed definition of the ``$\xi$-candidates" category satisfying the demands stated above is best constructed by a team of experimental experts within the ATLAS and CMS collaborations deeply familiar with their detectors. However, it is likely that such a definition meeting the demands of sensitivity and selectivity will have several key characteristics which we would like to discuss here. We will then make illustrative estimates of the utility of a $\xi$-candidates definitions based on these characteristics. 

We make use of MadGraph aMC@NLO\cite{madgraph} simulations to produce our signal events at leading order with the lagrangian of Eq.~\ref{eq:theory}, which are then hadronized via pythia 8\cite{pythia1,pythia2}. For our detector studies we utilize Delphes\cite{delphes} fast detector simulation framework with the default CMS card and FastJet\cite{fastjet} for jet clustering algorithms. 

To begin one must have a cluster, established by standard techniques. One useful criteria to impose on the pre-$\xi$-candidate cluster is a strong isolation requirement against QCD activity within a small cone around the $\xi$-candidate system, reducing QCD backgrounds from decaying pions. Additional criteria for the definition must also appeal to the stoutness of the photon jet --- there are two photons separated enough to not look like one photon and that separation shows up as a larger-than-normal spatial spread among cells within the electromagnetic detector.  Furthermore, vetoing on charged tracks eliminates electron-induced showers. Finally, recently established jet n-subjettiness algorithms\cite{Thaler:2010tr} can be employed to select clusters that have discernible sub-jet structure compatible with 2 collimated photons. Refs.\cite{photon_jets,chakraborty_framework_2018} go into detail on the ability to use these and other, similar variables to separate $\xi$-candidates (called photon-jets in these papers) from photons and QCD jets, but all of these considerations will be in play in the definitions below.

Our $\xi$-jet theory definition was for underlying two-photon clusters with $\Delta R$ separation in the range of $0.04$ to $0.4$. In addition, within the range of $0.025<\Delta R<0.04$ there is a possibility of using electromagnetic shape variables to discern that the underlying event was likely not a single photon, but certainly not clear enough to indicate the possibility of two photons. Nevertheless, our $\xi$-candidate list of criteria will be applicable for two-photon jets separations down to about $\Delta R\gsim 0.025$ and up to about $\Delta R \lsim 0.25$. We will not discuss the range $0.25\lsim \Delta R\lsim 0.4$ here, because our understanding is that more traditional photon identification tools may be applicable to separate the photons just well enough to help discern signal from photon backgrounds.

Let us now turn to a more precise definition of $\xi$-candidates (underlying two-photon separation $0.025\lsim \Delta R\lsim 0.25$). This regime targets events that have two photons in sufficiently close proximity that their cores overlap, thereby interfering with one anothers' identification procedure. This should appear as a cluster of energy in the EM calorimeter, with no tracks or corresponding energy in the hadronic calorimeter, and high 2-subjettiness. We provide an example definition of $\xi$-candidate criteria in Table~\ref{tab:xi}. Below in Fig.~\ref{fig:discr} we also show distributions of signal and background for QCD jets and $\xi$-jets. These distributions reproduce those of \cite{chakraborty_framework_2018,photon_jets} and show that $\xi$-jets can be separated from QCD backgrounds with high efficiency. 

\begin{table}[t]
    \centering
 \resizebox{\textwidth}{!}{%
   \begin{tabular}{c|c|c|c}
    \hline\hline
        Variable & Definition & Cut & Reasoning\\ \hline
        log$\theta_J$ & hadronic energy fraction & $< -0.8$ & exclude QCD and $\tau$ \\ \hline
        $N_T$ & Number of tracks & = 0 & excludes single converted\\
        & & & photons and jet activity \\ \hline
        $\tau_2/\tau_1$ & Ratio of 2- to 1-subjettiness & $< 0.3$ & Selects events with 2 subjets\\ \hline
            \end{tabular}}
    \caption{$\xi$ definition meant to capture underlying events with for events with $0.025 < \Delta R < 0.25$. These objects are defined as a cone of radius $\Delta R = 0.25$ about a central cluster in the EM calorimeter, centered on the highest energy pixel. Unless otherwise stated, the region is within the $\xi$ region. Here $\theta_{J}$ is the hadronic energy fraction.}
    \label{tab:xi}
\end{table} 

\begin{figure}[t]
\begin{center}
\subfloat[]{{\includegraphics[width=7cm]{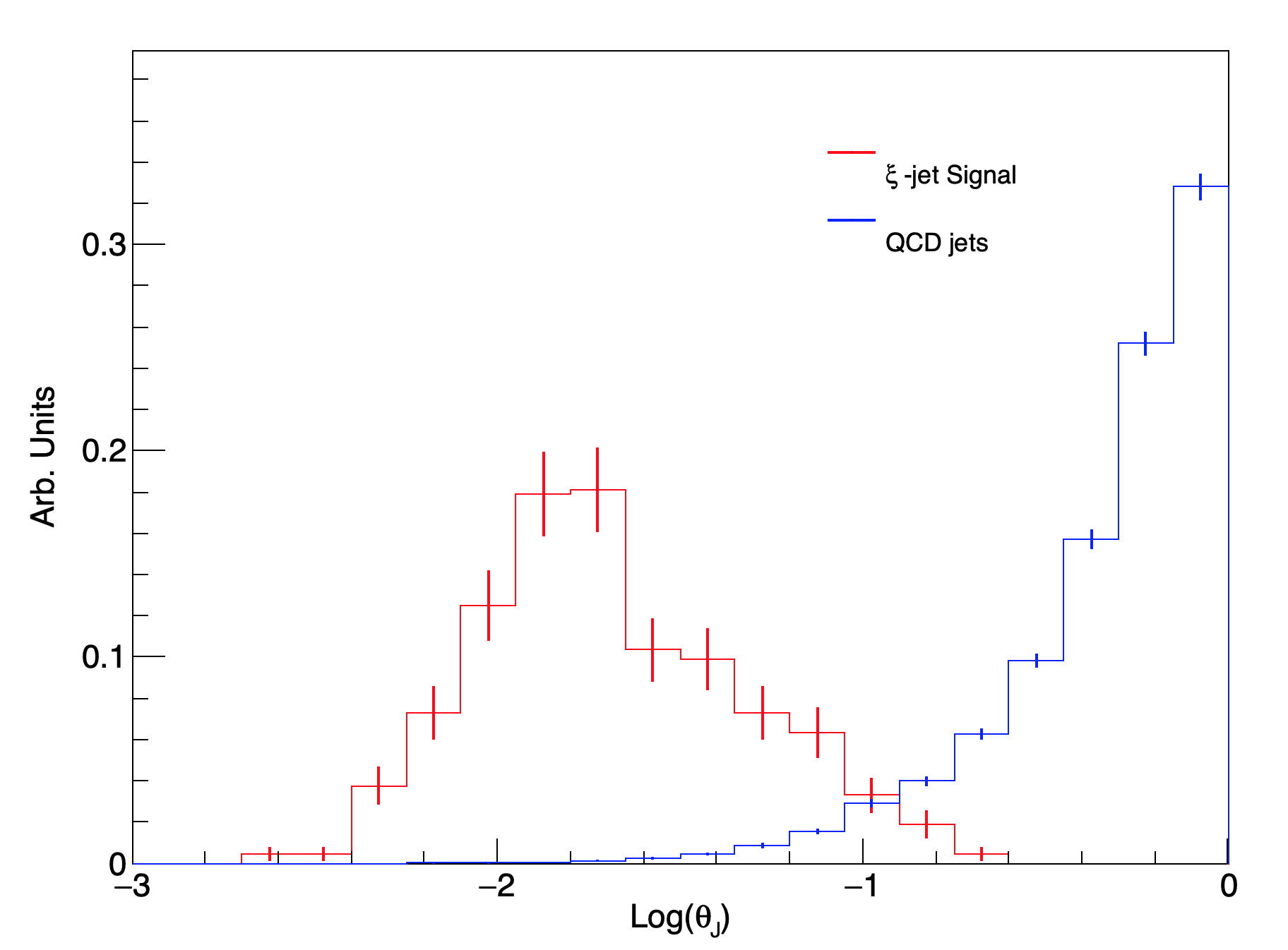} }}
\qquad
\subfloat[]{{\includegraphics[width=7cm]{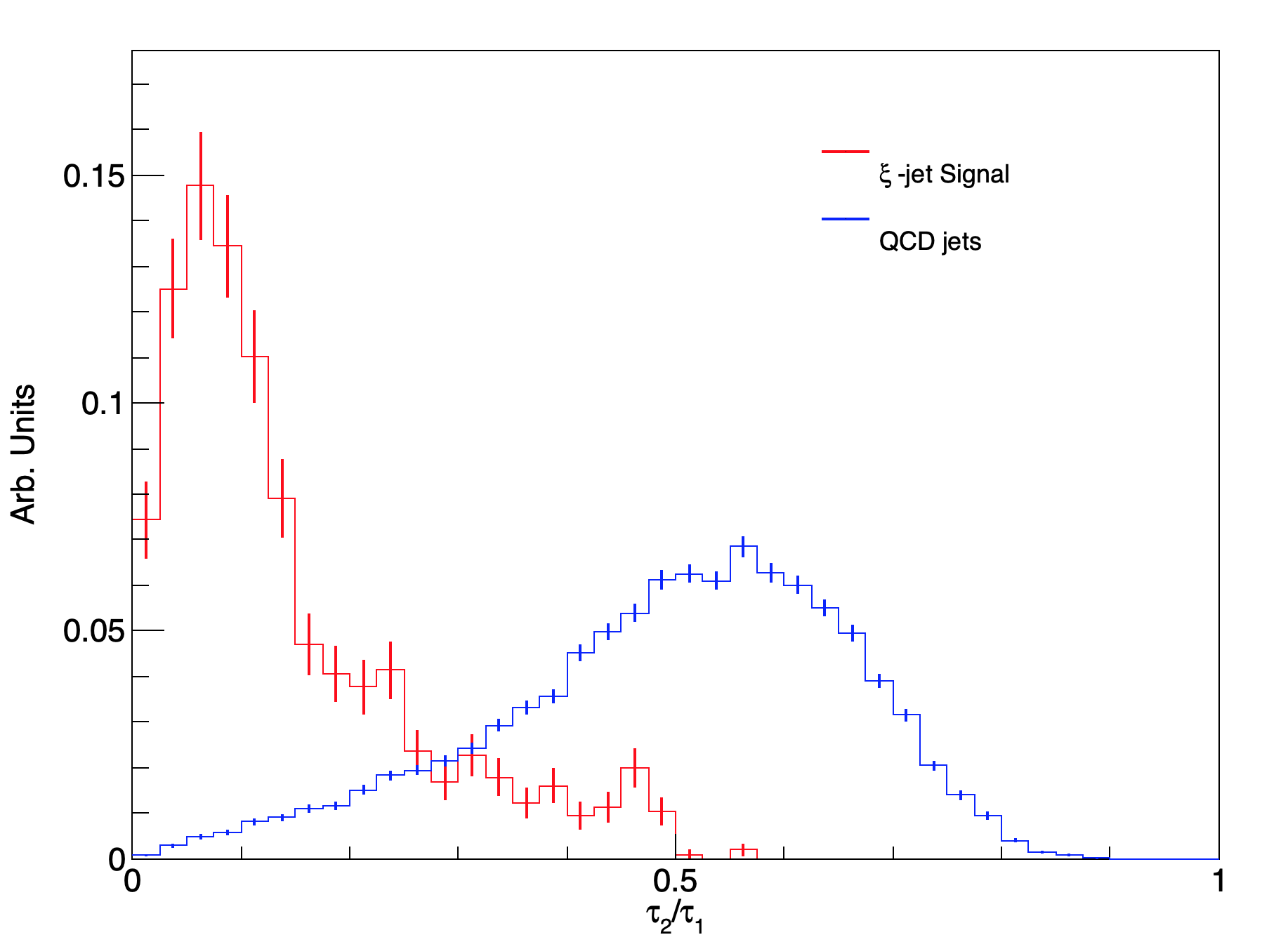} }}
\caption{Subset of kinematic variables useful for discriminating of $\xi$-jets (green) and QCD jets (blue) which are a major background. Here $\theta_{J}$ is the hadronic energy fraction for a jet, and $\tau_{2}/\tau_{1}$ is the ratio of 2-jettiness to 1-jettiness which is useful for picking out events with 2 subjets.}
\label{fig:discr}
\end{center}
\end{figure}

\subsection{Reconstructing $\xi$-jets and Higgs decays}

\begin{figure}[htbp]
\begin{center}
\includegraphics[width=0.85\textwidth]{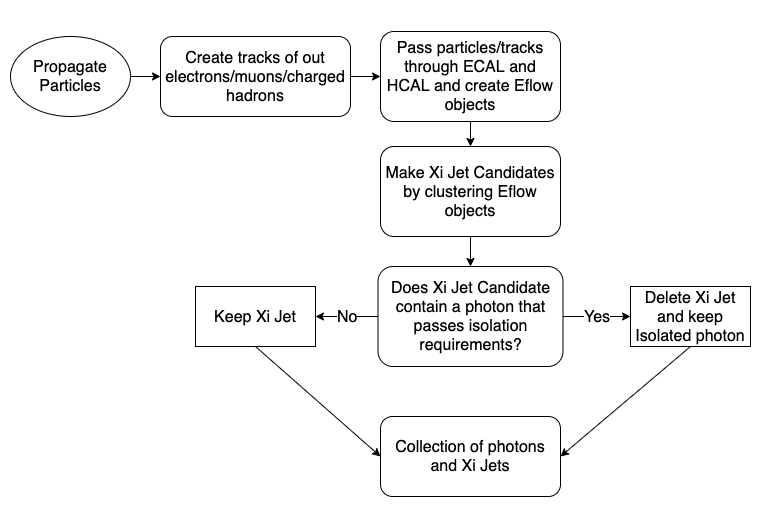}
\caption{the analysis flow of our reconstruction of $\xi$-jets using Delphes fast detector simulation. Additional photons not covered by that flow, as well as electrons, muons, jets, etc.\ are identified and labeled by other analysis flows. }
\label{fig:flowchart}
\end{center}
\end{figure}

Now that we have precise definitions of photons and $\xi$-candidates we can ask how well the Higgs boson signal can be reconstructed, especially in the case of its decay into one or more $\xi$-jets. Fig.~\ref{fig:flowchart} shows the analysis flow of our reconstruction of $\xi$-jets using Delphes fast detector simulation. Additional photons not covered by that flow, as well as electrons, muons, jets, etc.\ are identified and labeled by other analysis flows.

First, one must reconstruct the $\xi$-jets which we attempt to do by following a strategy similar to Ref.~\cite{photon_jets}. The method is as follows. First energy flow (eflow) objects\cite{particleflow} (composed of deposits in calorimeter cells) are clustered into jets using the anti-kt algorithm with $R = 0.25$. Then we re-cluster those energy deposits that were found in each jet using the kt algorithm, which determines a recombination tree for the jets. This tree specifies the subjets at each level of recombination $N$ from $N = 1$ (the full jet) to $N =$ the number of constituent eflow objects in the jet (no recombination). From here we can compute the $N$-subjettiness variable for the jet for each $N$. This variable becomes small when the parameter N is large enough to describe all of the relevant substructure of the jet. It is defined to be
\begin{equation}
\tau_{N} = \frac{\Sigma_{k} p_{T_{k}} \times {\rm min}[\Delta R_{1,k},\Delta R_{2,k}, . . . , \Delta R_{N,k}]}{\Sigma_{k}p_{T_{k}} \times R},
\end{equation}
where $k$ runs over all the constituents of the jet, $p_{T_{k}}$ is the transverse momentum for the $k$-th constituent, and $R$ is the characteristic jet radius used in the original jet clustering algorithm. 

After jet clustering is completed we then check if a reconstructed $\xi$-candidate already contains a reconstructed photon. Reconstructed photons are composed of eflow objects originating from the ECAL which must pass isolation requirements (cuts on electromagnetic and hadronic activity within a cone around the photon). If a $\xi$-candidate contains an already reconstructed, isolated photon then this $\xi$-candidate is deleted. 

Before applying additional cuts, we would like to characterize the efficiency at which we reconstruct $\xi$-jets. To do this we utilize Delphes GenJet objects. GenJets are jets that are clustered, not with calorimeter cells or towers or eflow objects, but with the actual generator level particles. By utilizing GenJets we can define ``generated $\xi$-jets" and see at what rate we correctly reconstruct these.

GenJets are clustered with the same strategy as above, first with the anti-kt algorithm with $R = 0.4$, and then reclustered with the kt algorithm. A GenJet is selected as a generated $\xi$-jet if it has: 1) At most two photons with  $p_T>$ 0.5 GeV, 2) no non-photons with $p_T >$ 0.5 GeV. Since our theoretical $\xi$-jets were defined as pairs of photons with $\Delta R$ between 0.04 and 0.4, we throw out $\xi$-jets with $\Delta R < 0.025$ as these will most likely be reconstructed as one photon. 

Once a generated $\xi$-jet is identified, we loop over all reconstructed $\xi$-candidates and attempt to find a match. Matching is done by comparing the $\Delta R$ between the momentum of the generated and reconstructed jets. If $\Delta R_{gen/reco} < 0.05$ we consider this jet as matched. We also require that the reconstructed $\xi$-jets pass a cut on the required hadronic energy fraction. This cut is that $\log(E_{had}/E_{jet}) < -0.8$.  Below in Fig.~\ref{fig:gen_reco} we show $\Delta R_{gen/reco}$, which shows the level of matching between generated and reconstructed $\xi$-jets. It also serves as a check that this is independent of our model parameters. 
\begin{figure}[t]
\begin{center}
\includegraphics[width=0.85\textwidth]{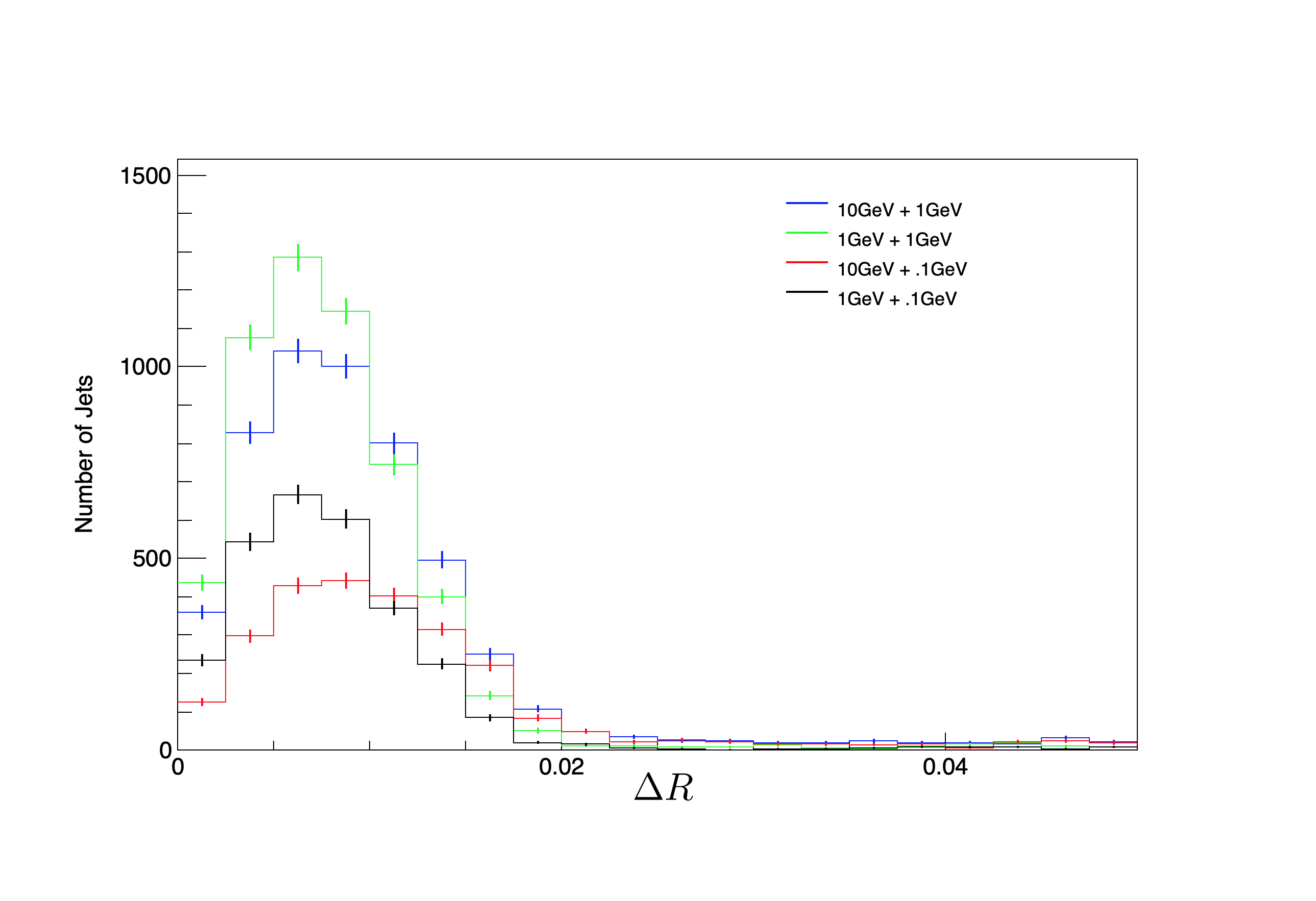}

\caption{$\Delta R$ between reconstructed and generated $\xi$-jets. Distribution is independent of our model parameters showing good matching between the two.}
\label{fig:gen_reco}
\end{center}
\end{figure}

Now we would like to understand how often we can reconstruct the Higgs mass using our reconstructed photons and $\xi$-candidates. To simplify matters we will choose $m_{\phi_{1}} = m_{\phi_{2}}$, which is equivalent to having only one light scalar in addition to the observed $h_{125}$. We scan over light scalar masses from 100 MeV to 14 GeV. This range ensures we see a smooth transition between photon dominated decays and $\xi$-jet dominated decays. The following discussion can be generalized by choosing different masses for the light scalars. After reconstruction, we first collect all of our reconstructed objects, which for now are photons and $\xi$-candidates. We only require our reconstructed $\xi$-candidates to pass our hadronic energy fraction cut, otherwise no cuts (besides minimum $p_T$ cuts which are used for clustering). We then form all the possible subsets of this collection, which have between 1 and 4 objects (as at most the Higgs decayed into 4 separable photons).  If one combination of $\xi$-candidates and photons yields an invariant mass within a 3 GeV window around 125 GeV (122 GeV $< M_{inv} <$ 128 GeV) then we consider this a match. Virtually no events contain multiple combinations of photons and $\xi$-jets which satisfy this requirement. We split each match into the following categories based on what number and type of objects make up the matching set.

\begin{enumerate}
\item Photons only: Matches with 2, 3, or 4 photons
\item Photons + $\xi$-jets: Matches with 1 photon + 1 $\xi$, 2 photons + 1 $\xi$, 1 photon + 2 $\xi$
\item $\xi$-jets only: Events with 2 and only 2 $\xi$-jets
\item $\xi$-jets inclusive: Includes the Photons + $\xi$-jets category as well as the $\xi$-jets only category
\item Other combinations: Any combination not included above
\item All: Any match in the accepted mass range
\end{enumerate}

\begin{figure}[t!]
\begin{center}
\includegraphics[width=0.85\textwidth]{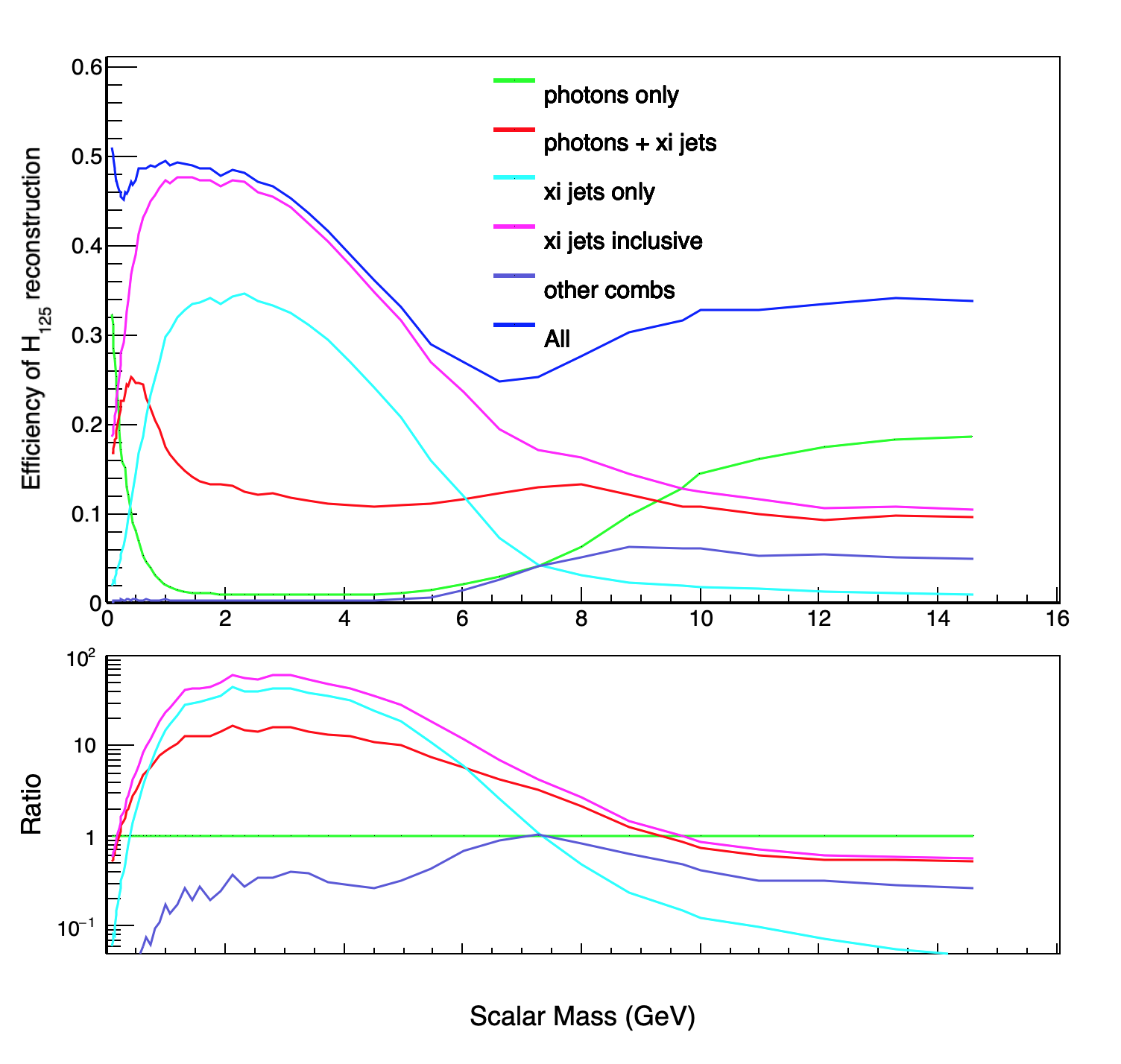}
\caption{Top: Efficiency of $h_{125}$ reconstruction as a function of scalar mass split into different categories based on number of $\xi$-jets and photons. Bottom: Ratio to only using photons. The reconstruction efficiency is more than an order of magnitude better when including $\xi$-jets (pink) over a wide range of masses from 100 MeV to 10 GeV. More specifically, the 2 $\xi$-jet channel dominates from the range of 300 MeV to 6 GeV.}
\label{fig:higgs_eff}
\end{center}
\end{figure}

Fig.~\ref{fig:higgs_eff} then shows the efficiency of reconstructing the Higgs mass as a function of the light scalar mass. Several key observations can be made here. At very low scalar masses, photons only makes up the dominant signal channel as the pairs of photons from $\phi$ decay are extremely collimated. From 100-300 MeV the signal from photons + $\xi$ jets becomes the most efficient channel as one of the pairs of photons is collimated enough to form a $\xi$-jet. Immediately above 300 MeV the signal from pairs of $\xi$-jets ($\xi$-jets only) becomes an order of magnitude more efficient than the photon only channel and remains so until ~6 GeV. Overall, searches including $\xi$-jets are more than an order of magnitude more efficient at reconstructing the Higgs from masses between 100 MeV and 10 GeV. 

Fig.~\ref{fig:higgs_eff} shows that searches including $\xi$-jets would be invaluable if a light scalar connected to the gauge and Higgs sector as in Eq.~\ref{eq:theory} exists in nature. We would like to stress that even though our analysis and definitions are quite simple, our results should be robust even after the introduction of more strict experimental search strategies and analysis cuts. It is interesting to compute how many such $\xi$-jet events one could expect for a given luminosity at the LHC. This is of course a function of the $\phi$ mass and the efficiency for reconstructing the $h_{125}$. To give an estimate, we can take the $m_{\phi} =$ 2 GeV point as an example. This has an efficiency for reconstruction of about 50\%. If we take Br($h\to\phi\phi$) = $10^{-4}$ and an integrated luminosity of $300\text{ fb}^{-1}$, this leaves us with about 7500 reconstructed events.

While a comprehensive study of standard model backgrounds is necessary for an experimental search, we can still make qualitative statements about discriminating $\xi$-jets from objects which fake $\xi$-jets. The largest backgrounds will be from $jj$, and $\gamma\, j$ production where a QCD jet fakes a $\xi$-jet. References~\cite{chakraborty_framework_2018,photon_jets} use a Boosted Decision Trees based on energy and substructure variables to discriminate between QCD jets and photon jets. They quote a fake rate from QCD jets of $10^{-4} - 10^{-5}$, though this fake rate is dependent on the rate at which one accidentally rejects $\xi$-jets. Additionally, the requirement that the invariant mass of the two $\xi$-candidates needs to fall within a 3 GeV window of the $h_{125}$ mass lowers the background as well, as the rate for QCD jet production tends to fall at high invariant masses. Combined, these factors should allow for a bump hunt search for $\xi$-jets with high sensitivity.

\section{Conclusion}

The discovery of the Higgs boson has lent strong support to the Standard Model, but also has allowed us to search for new avenues along which to extend it. In this work we have investigated exotic decays of the 125 GeV Higgs boson into light scalars which as of yet may be missed via current analysis techniques. We have discussed, first theoretically and then experimentally, a new object dubbed a $\xi$-jet which could play a pivotal role in the discovery of any light scalars minimally coupled to the standard model Higgs and to photons as in Eq.~\ref{eq:theory}. If experimentalists are able to identify and reconstruct $\xi$-jets these new objects could be strong evidence for an extended Higgs sector and Beyond the Standard Model physics. 

\section{Acknowledgments}  
We thank Advanced Research Computing at the University of Michigan, Ann Arbor for their computational resources, as well as D.~Amidei, C.~Hayes, R.~Hyneman and T.~Schwarz for helpful conversations on these issues. This work was supported in part by the DOE under grant DE-SC0007859. B. Sheff is supported by the NSF GRFP program, and N. Steinberg is supported by a fellowship from the Leinweber Center for Theoretical Physics.

\bibliographystyle{utphys}
\bibliography{xijets}

\end{document}